\documentclass{article} 
\usepackage{iclr2025_conference,times}

\usepackage{hyperref}
\usepackage{url}

\usepackage{latexsym}
\usepackage{multirow}
\usepackage{subfigure}
\usepackage{amsmath}
\usepackage{booktabs}
\usepackage{amssymb}
\usepackage{microtype}
\usepackage{balance}
\usepackage{inconsolata}
\usepackage{adjustbox}
\usepackage{graphicx}

\usepackage[utf8]{inputenc} 
\usepackage[T1]{fontenc}    
\usepackage{hyperref}       
\usepackage{url}            
\usepackage{booktabs}       
\usepackage{amsfonts}       
\usepackage{nicefrac}       
\usepackage{microtype}      
\usepackage{xcolor}         

\usepackage{multirow}
\usepackage{subfigure}

\usepackage[english]{babel}
\usepackage{moresize}
\usepackage{amsmath}
\usepackage{algorithmic}
\usepackage{balance}
\usepackage{comment}
\usepackage{paralist}
\usepackage{bm}
\usepackage{pgfplots}
\usetikzlibrary{pgfplots.dateplot}

\usepackage{flushend}
\usepackage[english]{babel}
\usepackage{graphicx}

\usepackage{amssymb}
\usepackage{amsfonts}
\usepackage{url}
\usepackage{bbm}
\usepackage{longtable}
\usepackage{rotating}
\usepackage{multirow}
\usepackage{mathrsfs}
\usepackage{enumitem}
\usepackage[linesnumbered,algoruled,boxed,lined]{algorithm2e}
\usepackage{adjustbox}
\usepackage{hyperref}
\usepackage{pgfplots}
\usetikzlibrary{pgfplots.dateplot}
\usepackage{filecontents}
\usepackage{wrapfig}
\usepackage{caption}

\def\model{LightRAG}

\title{LightRAG: Simple and Fast \\ Retrieval-Augmented Generation}

\author{Zirui Guo$^{1,2}$, Lianghao Xia$^{2}$, \textbf{Yanhua Yu}$^{1,*}$, Tu Ao$^{1}$, \textbf{Chao Huang}$^{2}$\thanks{Both Chao Huang and Yanhua Yu are corresponding authors.} \\
Beijing University of Posts and Telecommunications$^1$ \\
University of Hong Kong$^2$ \\
\texttt{\small zrguo101@hku.hk\; aka\_xia@foxmail.com\; chaohuang75@gmail.com}
}

\begin{document}

\maketitle

\begin{abstract}
Retrieval-Augmented Generation (RAG) systems enhance large language models (LLMs) by integrating external knowledge sources, enabling more accurate and contextually relevant responses tailored to user needs. However, existing RAG systems have significant limitations, including reliance on flat data representations and inadequate contextual awareness, which can lead to fragmented answers that fail to capture complex inter-dependencies. To address these challenges, we propose \model, which incorporates graph structures into text indexing and retrieval processes. This innovative framework employs a dual-level retrieval system that enhances comprehensive information retrieval from both low-level and high-level knowledge discovery. Additionally, the integration of graph structures with vector representations facilitates efficient retrieval of related entities and their relationships, significantly improving response times while maintaining contextual relevance. This capability is further enhanced by an incremental update algorithm that ensures the timely integration of new data, allowing the system to remain effective and responsive in rapidly changing data environments. Extensive experimental validation demonstrates considerable improvements in retrieval accuracy and efficiency compared to existing approaches. We have made our \model\ open-source and available at the link: \textcolor{blue}{\url{https://github.com/HKUDS/LightRAG}}.
\end{abstract}

\section{Introduction}
\label{sec:intro}

Retrieval-Augmented Generation (RAG) systems have been developed to enhance large language models (LLMs) by integrating external knowledge sources~\cite{sudhi2024rag, es2024ragas, salemi2024evaluating}. This innovative integration allows LLMs to generate more accurate and contextually relevant responses, significantly improving their utility in real-world applications. By adapting to specific domain knowledge~\cite{tu2024r}, RAG systems ensure that the information provided is not only pertinent but also tailored to the user's needs. Furthermore, they offer access to up-to-date information~\cite{zhao2024retrieval}, which is crucial in rapidly evolving fields. Chunking plays a vital role in facilitating the retrieval-augmented generation process~\cite{lyu2024crud}. By breaking down a large external text corpus into smaller, more manageable segments, chunking significantly enhances the accuracy of information retrieval. This approach allows for more targeted similarity searches, ensuring that the retrieved content is directly relevant to user queries.

However, existing RAG systems have key limitations that hinder their performance. \textbf{First}, many methods rely on flat data representations, restricting their ability to understand and retrieve information based on intricate relationships between entities. \textbf{Second}, these systems often lack the contextual awareness needed to maintain coherence across various entities and their interrelations, resulting in responses that may not fully address user queries. For example, consider a user asking, ``How does the rise of electric vehicles influence urban air quality and public transportation infrastructure?'' Existing RAG methods might retrieve separate documents on electric vehicles, air pollution, and public transportation challenges but struggle to synthesize this information into a cohesive response. They may fail to explain how the adoption of electric vehicles can improve air quality, which in turn could affect public transportation planning. As a result, the user may receive a fragmented answer that does not adequately capture the complex inter-dependencies among these topics.

To address these limitations, we propose incorporating graph structures into text indexing and relevant information retrieval. Graphs are particularly effective at representing the interdependencies among different entities~\cite{rampavsek2022recipe}, which enables a more nuanced understanding of relationships. The integration of graph-based knowledge structures facilitates the synthesis of information from multiple sources into coherent and contextually rich responses. Despite these advantages, developing a fast and scalable graph-empowered RAG system that efficiently handles varying query volumes is crucial. In this work, we achieve an effective and efficient RAG system by addressing three key challenges: i) \textbf{Comprehensive Information Retrieval}. Ensuring comprehensive information retrieval that captures the full context of inter-dependent entities from all documents; ii) \textbf{Enhanced Retrieval Efficiency}. Improving retrieval efficiency over the graph-based knowledge structures to significantly reduce response times; iii) \textbf{Rapid Adaptation to New Data}. Enabling quick adaptation to new data updates, ensuring the system remains relevant in dynamic environments.

In response to the outlined challenges, we propose \model, a model that seamlessly integrates a graph-based text indexing paradigm with a dual-level retrieval framework. This innovative approach enhances the system's capacity to capture complex inter-dependencies among entities, resulting in more coherent and contextually rich responses. \model\ employs efficient dual-level retrieval strategies: low-level retrieval, which focuses on precise information about specific entities and their relationships, and high-level retrieval, which encompasses broader topics and themes. By combining both detailed and conceptual retrieval, \model\ effectively accommodates a diverse range of quries, ensuring that users receive relevant and comprehensive responses tailored to their specific needs. Additionally, by integrating graph structures with vector representations, our framework facilitates efficient retrieval of related entities and relations while enhancing the comprehensiveness of results through relevant structural information from the constructed knowledge graph.

In summary, the key contributions of this work are highlighted as follows:

\begin{itemize}[leftmargin=*]

\item \textbf{General Aspect}. We emphasize the importance of developing a graph-empowered RAG system to overcome the limitations of existing methods. By integrating graph structures into text indexing, we can effectively represent complex interdependencies among entities, fostering a nuanced understanding of relationships and enabling coherent, contextually rich responses.

\item \textbf{Methodologies}. To enable an efficient and adaptive RAG system, we propose \model, which integrates a dual-level retrieval paradigm with graph-enhanced text indexing. This approach captures both low-level and high-level information for comprehensive, cost-effective retrieval. By eliminating the need to rebuild the entire index, \model\ reduces computational costs and accelerates adaptation, while its incremental update algorithm ensures timely integration of new data, maintaining effectiveness in dynamic environments.

\item \textbf{Experimental Findings}. Extensive experiments were conducted to evaluate the effectiveness of \model\ in comparison to existing RAG models. These assessments focused on several key dimensions, including retrieval accuracy, model ablation, response efficiency, and adaptability to new information. The results demonstrated significant improvements over baseline methods.

\end{itemize}
\section{Retrieval-Augmented Generation}
\label{sec:model}
Retrieval-Augmented Generation (RAG) integrates user queries with a collection of pertinent documents sourced from an external knowledge database, incorporating two essential elements: the \textbf{Retrieval Component} and the \textbf{Generation Component}. 1) The retrieval component is responsible for fetching relevant documents or information from the external knowledge database. It identifies and retrieves the most pertinent data based on the input query. 2) After the retrieval process, the generation component takes the retrieved information and generates coherent, contextually relevant responses. It leverages the capabilities of the language model to produce meaningful outputs.
Formally, this RAG framework, denoted as $\mathcal{M}$, can be defined as follows:
\begin{align}
    \mathcal{M} = \Big( \mathcal{G}, ~~ \mathcal{R}=(\varphi, \psi) \Big),~~~
    \mathcal{M}(q; \mathcal{D}) = \mathcal{G}\Big(q, \psi(q; \hat{\mathcal{D}})\Big),~~~\hat{\mathcal{D}} = \varphi(\mathcal{D})
\end{align}
In this framework, $\mathcal{G}$ and $\mathcal{R}$ represent the generation module and the retrieval module, respectively, while $q$ denotes the input query and $D$ refers to the external database. The retrieval module $\mathcal{R}$ includes two key functionalities: i) \textbf{Data Indexer} $\varphi (\cdot)$: which involves building a specific data structure $\hat{\mathcal{D}}$ based on the external database $D$. ii) \textbf{Data Retriever} $\psi(\cdot)$: The relevant documents are obtained by comparing the query against the indexed data, also denoted as ``relevant documents''. By leveraging the information retrieved through $\psi(\cdot)$ along with the initial query $q$, the generative model $\mathcal{G}(\cdot)$ efficiently produces high-quality, contextually relevant responses.



In this work, we target several key points essential for an efficient and effective Retrieval-Augmented Generation (RAG) system which are elaborated below:\vspace{-0.05in}
\begin{itemize}[leftmargin=*]

\item \textbf{Comprehensive Information Retrieval}: The indexing function $\varphi(\cdot)$ must be adept at extracting global information, as this is crucial for enhancing the model's ability to answer queries effectively.

\item \textbf{Efficient and Low-Cost Retrieval}: The indexed data structure $\hat{\mathcal{D}}$ must enable rapid and cost-efficient retrieval to effectively handle a high volume of queries.

\item \textbf{Fast Adaptation to Data Changes}: The ability to swiftly and efficiently adjust the data structure to incorporate new information from the external knowledge base, is crucial for ensuring that the system remains current and relevant in an ever-changing information landscape.

\end{itemize}

\vspace{-0.1in}
\section{The \model\ Architecture}
\label{sec:solution}


\begin{figure}
    \centering
    \includegraphics[width=\textwidth]{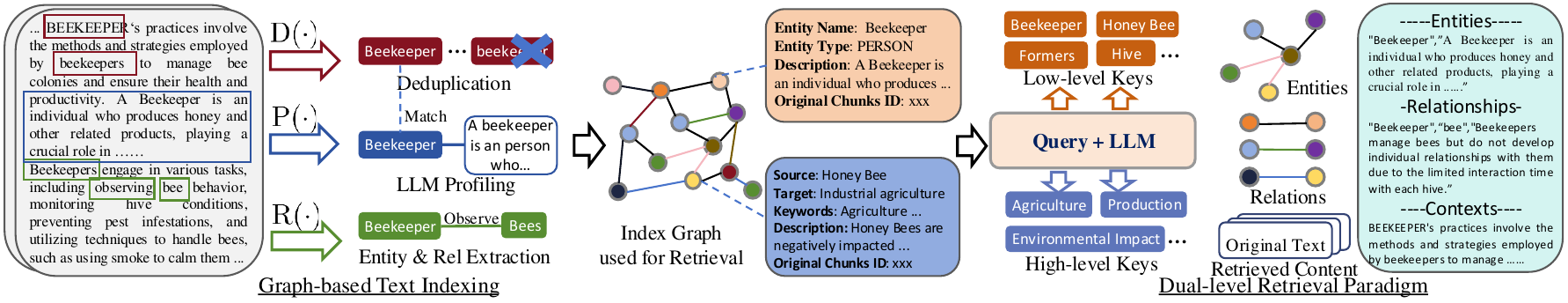}
    \vspace{-0.2in}
    \caption{Overall architecture of the proposed \model\ framework.}
    \label{fig:framework}
    \vspace{-0.1in}
\end{figure}

\subsection{Graph-based Text Indexing}

\textbf{Graph-Enhanced Entity and Relationship Extraction}. Our \model\ enhances the retrieval system by segmenting documents into smaller, more manageable pieces. This strategy allows for quick identification and access to relevant information without analyzing entire documents. Next, we leverage LLMs to identify and extract various entities (e.g., names, dates, locations, and events) along with the relationships between them. The information collected through this process will be used to create a comprehensive knowledge graph that highlights the connections and insights across the entire collection of documents. We formally represent this graph generation module as follows:
\begin{align}
    \hat{\mathcal{D}} = (\hat{\mathcal{V}},\hat{\mathcal{E}}) = \text{Dedupe}\circ\text{Prof}(\mathcal{V}, \mathcal{E}),~~~\mathcal{V}, \mathcal{E} = \cup_{\mathcal{D}_i\in\mathcal{D}} \text{Recog}(\mathcal{D}_i)
\end{align}
where $\hat{\mathcal{D}}$ represents the resulting knowledge graphs. To generate this data, we apply three main processing steps to the raw text documents $\mathcal{D}_i$. These steps utilize a LLM for text analysis and processing. Details about the prompt templates and specific settings for this part can be found in Appendix~\ref{app:graph_construct_prompt}. The functions used in our graph-based text indexing paradigm are described as:

\begin{itemize}[leftmargin=*]


    \item \textbf{Extracting Entities and Relationships}. $\text{R}(\cdot)$: This function prompts a LLM to identify entities (nodes) and their relationships (edges) within the text data. For instance, it can extract entities like "Cardiologists" and "Heart Disease," and relationships such as "Cardiologists diagnose Heart Disease" from the text: "Cardiologists assess symptoms to identify potential heart issues." To improve efficiency, the raw text $\mathcal{D}$ is segmented into multiple chunks $\mathcal{D}_i$.
    

    \item \textbf{LLM Profiling for Key-Value Pair Generation}. $\text{P}(\cdot)$: We employ a LLM-empowered profiling function, $\text{P}(\cdot)$, to generate a text key-value pair $(K, V)$ for each entity node in $\mathcal{V}$ and relation edge in $\mathcal{E}$. Each index key is a word or short phrase that enables efficient retrieval, while the corresponding value is a text paragraph summarizing relevant snippets from external data to aid in text generation. Entities use their names as the sole index key, whereas relations may have multiple index keys derived from LLM enhancements that include global themes from connected entities.
    

    \item \textbf{Deduplication to Optimize Graph Operations}. $\text{D}(\cdot)$: Finally, we implement a deduplication function, $\text{D}(\cdot)$, that identifies and merges identical entities and relations from different segments of the raw text $\mathcal{D}_i$. This process effectively reduces the overhead associated with graph operations on $\hat{\mathcal{D}}$ by minimizing the graph's size, leading to more efficient data processing.
    
\end{itemize}

Our \model\ offers two advantages through its graph-based text indexing paradigm. \emph{First}, \textbf{Comprehensive Information Understanding}. The constructed graph structures enable the extraction of global information from multi-hop subgraphs, greatly enhancing \model's ability to handle complex queries that span multiple document chunks. \emph{Second}, \textbf{Enhanced Retrieval Performance}. the key-value data structures derived from the graph are optimized for rapid and precise retrieval. This provides a superior alternative to less accurate embedding matching methods~\citep{gao2023retrieval} and inefficient chunk traversal techniques~\citep{edge2024local} commonly used in existing approaches.




\textbf{Fast Adaptation to Incremental Knowledge Base}. To efficiently adapt to evolving data changes while ensuring accurate and relevant responses, our \model\ incrementally updates the knowledge base without the need for complete reprocessing of the entire external database. For a new document $\mathcal{D}'$, the incremental update algorithm processes it using the same graph-based indexing steps $\varphi$ as before, resulting in $\hat{\mathcal{D}}' = (\hat{\mathcal{V}}', \hat{\mathcal{E}}')$. Subsequently, \model combines the new graph data with the original by taking the union of the node sets $\hat{\mathcal{V}}$ and $\hat{\mathcal{V}}'$, as well as the edge sets $\hat{\mathcal{E}}$ and $\hat{\mathcal{E}}'$.

Two key objectives guide our approach to fast adaptation for the incremental knowledge base: \textbf{Seamless Integration of New Data}. By applying a consistent methodology to new information, the incremental update module allows the \model\ to integrate new external databases without disrupting the existing graph structure. This approach preserves the integrity of established connections, ensuring that historical data remains accessible while enriching the graph without conflicts or redundancies. \textbf{Reducing Computational Overhead
}. By eliminating the need to rebuild the entire index graph, this method reduces computational overhead and facilitates the rapid assimilation of new data. Consequently, \model\ maintains system accuracy, provides current information, and conserves resources, ensuring users receive timely updates and enhancing the overall RAG effectiveness.


\subsection{Dual-level Retrieval Paradigm}
To retrieve relevant information from both specific document chunks and their complex inter-dependencies, our \model\ proposes generating query keys at both detailed and abstract levels.\vspace{-0.05in}
\begin{itemize}[leftmargin=*]

\item \textbf{Specific Queries}. These queries are detail-oriented and typically reference specific entities within the graph, requiring precise retrieval of information associated with particular nodes or edges. For example, a specific query might be, ``Who wrote 'Pride and Prejudice'?''

\item \textbf{Abstract Queries}. In contrast, abstract queries are more conceptual, encompassing broader topics, summaries, or overarching themes that are not directly tied to specific entities. An example of an abstract query is, ``How does artificial intelligence influence modern education?''

\end{itemize}


To accommodate diverse query types, the \model\ employs two distinct retrieval strategies within the dual-level retrieval paradigm. This ensures that both specific and abstract inquiries are addressed effectively, allowing the system to deliver relevant responses tailored to user needs.
\begin{itemize}[leftmargin=*]

\item \textbf{Low-Level Retrieval}. This level is primarily focused on retrieving specific entities along with their associated attributes or relationships. Queries at this level are detail-oriented and aim to extract precise information about particular nodes or edges within the graph.

\item \textbf{High-Level Retrieval}. This level addresses broader topics and overarching themes. Queries at this level aggregate information across multiple related entities and relationships, providing insights into higher-level concepts and summaries rather than specific details.

\end{itemize}

\textbf{Integrating Graph and Vectors for Efficient Retrieval}.
By combining graph structures with vector representations, the model gains a deeper insight into the interrelationships among entities. This synergy enables the retrieval algorithm to effectively utilize both local and global keywords, streamlining the search process and improving the relevance of results.\vspace{-0.05in}

\begin{itemize}[leftmargin=*]

\item (i) \textbf{Query Keyword Extraction}. For a given query $q$, the retrieval algorithm of \model\ begins by extracting both local query keywords $k^{(l)}$ and global query keywords $k^{(g)}$.

\item (ii) \textbf{Keyword Matching}. The algorithm uses an efficient vector database to match local query keywords with candidate entities and global query keywords with relations linked to global keys.

\item (iii) \textbf{Incorporating High-Order Relatedness}. To enhance the query with higher-order relatedness, \model further gathers neighboring nodes within the local subgraphs of the retrieved graph elements. This process involves the set $\{v_i | v_i \in \mathcal{V} \land (v_i \in \mathcal{N}_v \lor v_i \in \mathcal{N}_e)\}$, where $\mathcal{N}_v$ and $\mathcal{N}_e$ represent the one-hop neighboring nodes of the retrieved nodes $v$ and edges $e$, respectively.

\end{itemize}

This dual-level retrieval paradigm not only facilitates efficient retrieval of related entities and relations through keyword matching, but also enhances the comprehensiveness of results by integrating relevant structural information from the constructed knowledge graph.


\subsection{Retrieval-Augmented Answer Generation}
\textbf{Utilization of Retrieved Information}. Utilizing the retrieved information $\psi(q;\hat{\mathcal{D}})$, our \model\ employs a general-purpose LLM to generate answers based on the collected data. This data comprises concatenated values $V$ from relevant entities and relations, produced by the profiling function $\text{P}(\cdot)$. It includes names, descriptions of entities and relations, and excerpts from the original text.

\textbf{Context Integration and Answer Generation}. By unifying the query with this multi-source text, the LLM generates informative answers tailored to the user's needs, ensuring alignment with the query's intent. This approach streamlines the answer generation process by integrating both context and query into the LLM model, as illustrated in detailed examples (Appendix~\ref{app:retrieval_and_generation_examples}).

\subsection{Complexity Analysis of the \model\ Framework}
In this section, we analyze the complexity of our proposed LightRAG framework, which can be divided into two main parts. The first part is the graph-based Index phase. During this phase, we use the large language model (LLM) to extract entities and relationships from each chunk of text. As a result, the LLM needs to be called \(\frac{\text{total tokens}}{\text{chunk size}}\) times. Importantly, there is no additional overhead involved in this process, making our approach highly efficient in managing updates to new text.

The second part of the process involves the graph-based retrieval phase. For each query, we first utilize the large language model (LLM) to generate relevant keywords. Similar to current Retrieval-Augmented Generation (RAG) systems \cite{gao2023retrieval, hyde, chan2024rq}, our retrieval mechanism relies on vector-based search. However, instead of retrieving chunks as in conventional RAG, we concentrate on retrieving entities and relationships. This approach markedly reduces retrieval overhead compared to the community-based traversal method used in GraphRAG.





\section{Evaluation}
\label{sec:eval}
We conduct empirical evaluations on benchmark data to assess the effectiveness of the proposed \model\ framework by addressing the following research questions: $\bullet$ \textbf{(RQ1)}: How does \model\ compare to existing RAG baseline methods in terms of generation performance? $\bullet$ \textbf{(RQ2)}: How do dual-level retrieval and graph-based indexing enhance the generation quality of \model? $\bullet$ \textbf{(RQ3)}: What specific advantages does \model\ demonstrate through case examples in various scenarios? $\bullet$ \textbf{(RQ4)}: What are the costs associated with \model, as well as its adaptability to data changes?

\subsection{Experimental Settings}
\textbf{Evaluation Datasets}.
To conduct a comprehensive analysis of \model, we selected four datasets from the UltraDomain benchmark~\citep{qian2024memorag}. The UltraDomain data is sourced from 428 college textbooks and encompasses 18 distinct domains, including agriculture, social sciences, and humanities. From these, we chose the Agriculture, CS, Legal, and Mix datasets. Each dataset contains between 600,000 and 5,000,000 tokens, with detailed information provided in Table~\ref{dataset}. Below is a specific introduction to the four domains utilized in our experiments:
\begin{itemize}[leftmargin=*]
    \item \textbf{Agriculture}: This domain focuses on agricultural practices, covering a range of topics including beekeeping, hive management, crop production, and disease prevention.
    
    \item  \textbf{CS}: This domain focuses on computer science and encompasses key areas of data science and software engineering. It particularly highlights machine learning and big data processing, featuring content on recommendation systems, classification algorithms, and real-time analytics using Spark.
    
    \item \textbf{Legal}: This domain centers on corporate legal practices, addressing corporate restructuring, legal agreements, regulatory compliance, and governance, with a focus on the legal and financial sectors.
    
    \item \textbf{Mixed}: This domain presents a rich variety of literary, biographical, and philosophical texts, spanning a broad spectrum of disciplines, including cultural, historical, and philosophical studies.
\end{itemize}

\textbf{Question Generation}. 
To evaluate the effectiveness of RAG systems for high-level sensemaking tasks, we consolidate all text content from each dataset as context and adopt the generation method outlined in~\cite{edge2024local}. Specifically, we instruct an LLM to generate five RAG users, along with five tasks for each user. Each generated user is accompanied by a textual description detailing their expertise and traits that motivate their question-raising activities. Each user task is also described, emphasizing one of the user's potential intentions when interacting with RAG systems. For each user-task combination, the LLM generates five questions that require an understanding of the entire corpus. In total, this process results in 125 questions for each dataset.

\textbf{Baselines}.  \model\ is compared against the following state-of-the-art methods across all datasets:
\begin{itemize}[leftmargin=*]
    \item \textbf{Naive RAG}~\citep{gao2023retrieval}: This model serves as a standard baseline in existing RAG systems. It segments raw texts into chunks and stores them in a vector database using text embeddings. For queries, Naive RAG generates vectorized representations to directly retrieve text chunks based on the highest similarity in their representations, ensuring efficient and straightforward matching.
    
    \item \textbf{RQ-RAG}~\citep{chan2024rq}: This approach leverages the LLM to decompose the input query into multiple sub-queries. These sub-queries are designed to enhance search accuracy by utilizing explicit techniques such as rewriting, decomposition, and disambiguation.
    
    \item \textbf{HyDE}~\citep{hyde}: This method utilizes the LLM to generate a hypothetical document based on the input query. This generated document is then employed to retrieve relevant text chunks, which are subsequently used to formulate the final answer.
    
    \item \textbf{GraphRAG}~\citep{edge2024local}: This is a graph-enhanced RAG system that utilizes an LLM to extract entities and relationships from the text, representing them as nodes and edges. It generates corresponding descriptions for these elements, aggregates nodes into communities, and produces a community report to capture global information. When handling high-level queries, GraphRAG retrieves more comprehensive information by traversing these communities.
\end{itemize}

\textbf{Implementation and Evaluation Details}.
In our experiments, we utilize the \href{https://github.com/gusye1234/nano-vectordb}{\textit{nano vector database}} for vector data management and access. For all LLM-based operations in \model, we default to using GPT-4o-mini. To ensure consistency, the chunk size is set to 1200 across all datasets. Additionally, the gleaning parameter is fixed at 1 for both GraphRAG and \model.

Defining ground truth for many RAG queries, particularly those involving complex high-level semantics, poses significant challenges. To address this, we build on existing work~\citep{edge2024local} and adopt an LLM-based multi-dimensional comparison method. We employ a robust LLM, specifically GPT-4o-mini, to rank each baseline against our \model. The evaluation prompt we used is detailed in Appendix~\ref{app:evaluation_prompt}. In total, we utilize four evaluation dimensions, including:

i) \textbf{Comprehensiveness}: How thoroughly does the answer address all aspects and details of the question? ii) \textbf{Diversity}: How varied and rich is the answer in offering different perspectives and insights related to the question? iii) \textbf{Empowerment}: How effectively does the answer enable the reader to understand the topic and make informed judgments? iv) \textbf{Overall}: This dimension assesses the cumulative performance across the three preceding criteria to identify the best overall answer.

The LLM directly compares two answers for each dimension and selects the superior response for each criterion. After identifying the winning answer for the three dimensions, the LLM combines the results to determine the overall better answer. To ensure a fair evaluation and mitigate the potential bias that could arise from the order in which the answers are presented in the prompt, we alternate the placement of each answer. We calculate win rates accordingly, ultimately leading to the final results. 

\subsection{Comparison of \model\ with Existing RAG Methods (RQ1)}

We compare \model\ against each baseline across various evaluation dimensions and datasets. The results are presented in Table~\ref{tab:performance}. Based on these findings, we draw the following conclusions:

\begin{table}[t]
\centering
\caption{Win rates (\%) of baselines v.s. \model\ across four datasets and four evaluation dimensions.}
\label{tab:performance}
\vspace{-0.1in}
\resizebox{\textwidth}{!}{
\begin{tabular}{@{}lcccccccc@{}}
\toprule
\textbf{}    & \multicolumn{2}{c}{\textbf{Agriculture}} & \multicolumn{2}{c}{\textbf{CS}} & \multicolumn{2}{c}{\textbf{Legal}} & \multicolumn{2}{c}{\textbf{Mix}} \\ 
\cmidrule(lr){2-3} \cmidrule(lr){4-5} \cmidrule(lr){6-7} \cmidrule(lr){8-9}
                      & NaiveRAG & \textbf{LightRAG} & NaiveRAG & \textbf{LightRAG} & NaiveRAG & \textbf{LightRAG} & NaiveRAG & \textbf{LightRAG} \\
\midrule
Comprehensiveness      & 32.4\%      & \underline{67.6\%}     & 38.4\%      & \underline{61.6\%}     & 16.4\%      & \underline{83.6\%}     & 38.8\%      & \underline{61.2\%}     \\
Diversity              & 23.6\%      & \underline{76.4\%}     & 38.0\%      & \underline{62.0\%}     & 13.6\%      & \underline{86.4\%}     & 32.4\%      & \underline{67.6\%}     \\
Empowerment            & 32.4\%      & \underline{67.6\%}     & 38.8\%      & \underline{61.2\%}     & 16.4\%      & \underline{83.6\%}     & 42.8\%      & \underline{57.2\%}     \\
Overall                & 32.4\%      & \underline{67.6\%}     & 38.8\%      & \underline{61.2\%}     & 15.2\%      & \underline{84.8\%}     & 40.0\%      & \underline{60.0\%}     \\
\cmidrule(lr){2-3} \cmidrule(lr){4-5} \cmidrule(lr){6-7} \cmidrule(lr){8-9}
                      & RQ-RAG & \textbf{LightRAG} & RQ-RAG & \textbf{LightRAG} & RQ-RAG & \textbf{LightRAG} & RQ-RAG & \textbf{LightRAG} \\
\midrule
Comprehensiveness      & 31.6\%      & \underline{68.4\%}     & 38.8\%      & \underline{61.2\%}     & 15.2\%      & \underline{84.8\%}     & 39.2\%      & \underline{60.8\%}     \\
Diversity              & 29.2\%      & \underline{70.8\%}     & 39.2\%      & \underline{60.8\%}     & 11.6\%      & \underline{88.4\%}     & 30.8\%      & \underline{69.2\%}     \\
Empowerment            & 31.6\%      & \underline{68.4\%}     & 36.4\%      & \underline{63.6\%}     & 15.2\%      & \underline{84.8\%}     & 42.4\%      & \underline{57.6\%}     \\
Overall         & 32.4\%      & \underline{67.6\%}     & 38.0\%      & \underline{62.0\%}     & 14.4\%      & \underline{85.6\%}     & 40.0\%      & \underline{60.0\%}     \\
\cmidrule(lr){2-3} \cmidrule(lr){4-5} \cmidrule(lr){6-7} \cmidrule(lr){8-9}
                      & HyDE & \textbf{LightRAG} & HyDE & \textbf{LightRAG} & HyDE & \textbf{LightRAG} & HyDE & \textbf{LightRAG} \\
\midrule
Comprehensiveness      & 26.0\%      & \underline{74.0\%}     & 41.6\%      & \underline{58.4\%}     & 26.8\%      & \underline{73.2\%}     & 40.4\%      & \underline{59.6\%}     \\
Diversity              & 24.0\%      & \underline{76.0\%}     & 38.8\%      & \underline{61.2\%}     & 20.0\%      & \underline{80.0\%}     & 32.4\%      & \underline{67.6\%}     \\
Empowerment            & 25.2\%      & \underline{74.8\%}     & 40.8\%      & \underline{59.2\%}     & 26.0\%      & \underline{74.0\%}     & 46.0\%      & \underline{54.0\%}     \\
Overall         & 24.8\%      & \underline{75.2\%}     & 41.6\%      & \underline{58.4\%}     & 26.4\%      & \underline{73.6\%}     & 42.4\%      & \underline{57.6\%}     \\
\cmidrule(lr){2-3} \cmidrule(lr){4-5} \cmidrule(lr){6-7} \cmidrule(lr){8-9}
                      & GraphRAG & \textbf{LightRAG} & GraphRAG & \textbf{LightRAG} & GraphRAG & \textbf{LightRAG} & GraphRAG & \textbf{LightRAG} \\
\midrule
Comprehensiveness      & 45.6\%      & \underline{54.4\%}     & 48.4\%      & \underline{51.6\%}     &  48.4\%     & \underline{51.6\%}     & \underline{50.4\%}      & 49.6\%     \\
Diversity              & 22.8\%      & \underline{77.2\%}     & 40.8\%      & \underline{59.2\%}     &  26.4\%     & \underline{73.6\%}     & 36.0\%      & \underline{64.0\%}     \\
Empowerment            & 41.2\%      & \underline{58.8\%}     & 45.2\%      & \underline{54.8\%}     & 43.6\%      & \underline{56.4\%}     & \underline{50.8\%}      & 49.2\%     \\
Overall                & 45.2\%      & \underline{54.8\%}     & 48.0\%      & \underline{52.0\%}     & 47.2\%      & \underline{52.8\%}     & \underline{50.4\%}      & 49.6\%     \\
\bottomrule
\end{tabular}
}
\end{table}

\textbf{The Superiority of Graph-enhanced RAG Systems in Large-Scale Corpora}
When handling large token counts and complex queries that require a thorough understanding of the dataset's context, graph-based RAG systems like \model\ and GraphRAG consistently outperform purely chunk-based retrieval methods such as NaiveRAG, HyDE, and RQRAG. This performance gap becomes particularly pronounced as the dataset size increases. For instance, in the largest dataset (Legal), the disparity widens significantly, with baseline methods achieving only about 20\% win rates compared to the dominance of \model. This trend underscores the advantages of graph-enhanced RAG systems in capturing complex semantic dependencies within large-scale corpora, facilitating a more comprehensive understanding of knowledge and leading to improved generalization performance.

\textbf{Enhancing Response Diversity with \model}:
Compared to various baselines, \model\ demonstrates a significant advantage in the Diversity metric, particularly within the larger Legal dataset. Its consistent lead in this area underscores \model's effectiveness in generating a wider range of responses, especially in scenarios where diverse content is essential. We attribute this advantage to \model's dual-level retrieval paradigm, which facilitates comprehensive information retrieval from both low-level and high-level dimensions. This approach effectively leverages graph-based text indexing to consistently capture the full context in response to queries.

\textbf{\model's Superiority over GraphRAG}: 
While both \model\ and GraphRAG use graph-based retrieval mechanisms, \model\ consistently outperforms GraphRAG, particularly in larger datasets with complex language contexts. In the Agriculture, CS, and Legal datasets—each containing millions of tokens—\model\ shows a clear advantage, significantly surpassing GraphRAG and highlighting its strength in comprehensive information understanding within diverse environments. \textbf{Enhanced Response Variety}: By integrating low-level retrieval of specific entities with high-level retrieval of broader topics, \model\ boosts response diversity. This dual-level mechanism effectively addresses both detailed and abstract queries, ensuring a thorough grasp of information. \textbf{Complex Query Handling}: This approach is especially valuable in scenarios requiring diverse perspectives. By accessing both specific details and overarching themes, \model\ adeptly responds to complex queries involving interconnected topics, providing contextually relevant answers.

\subsection{Ablation Studies (RQ2)}
\begin{table}[t]
\centering
\caption{Performance of ablated versions of \model, using NaiveRAG as reference.}
\label{tab:ablation}
\vspace{-0.1in}
\setlength{\tabcolsep}{1.2\tabcolsep}
\resizebox{\textwidth}{!}{
\begin{tabular}{@{}lcccccccc@{}}
\toprule
\textbf{}    & \multicolumn{2}{c}{\textbf{Agriculture}} & \multicolumn{2}{c}{\textbf{CS}} & \multicolumn{2}{c}{\textbf{Legal}} & \multicolumn{2}{c}{\textbf{Mix}} \\ 
\cmidrule(lr){2-3} \cmidrule(lr){4-5} \cmidrule(lr){6-7} \cmidrule(lr){8-9}
                      & NaiveRAG & \textbf{\model} & NaiveRAG & \textbf{\model} & NaiveRAG & \textbf{\model} & NaiveRAG & \textbf{\model} \\
\midrule
Comprehensiveness      & 32.4\%      & \underline{67.6\%}     & 38.4\%      & \underline{61.6\%}     & 16.4\%      & \underline{83.6\%}     & 38.8\%      & \underline{61.2\%}     \\
Diversity              & 23.6\%      & \underline{76.4\%}     & 38.0\%      & \underline{62.0\%}     & 13.6\%      & \underline{86.4\%}     & 32.4\%      & \underline{67.6\%}     \\
Empowerment            & 32.4\%      & \underline{67.6\%}     & 38.8\%      & \underline{61.2\%}     & 16.4\%      & \underline{83.6\%}     & 42.8\%      & \underline{57.2\%}     \\
Overall                & 32.4\%      & \underline{67.6\%}     & 38.8\%      & \underline{61.2\%}     & 15.2\%      & \underline{84.8\%}     & 40.0\%      & \underline{60.0\%}     \\
\cmidrule(lr){2-3} \cmidrule(lr){4-5} \cmidrule(lr){6-7} \cmidrule(lr){8-9}
                      & NaiveRAG & \textbf{-High} & NaiveRAG & \textbf{-High} & NaiveRAG & \textbf{-High} & NaiveRAG & \textbf{-High} \\
\midrule
Comprehensiveness      & 34.8\%      & \underline{65.2\%}     & 42.8\%      & \underline{57.2\%}     & 23.6\%      & \underline{76.4\%}     & 40.4\%      & \underline{59.6\%}     \\
Diversity              & 27.2\%      & \underline{72.8\%}     & 36.8\%      & \underline{63.2\%}     & 16.8\%      & \underline{83.2\%}     & 36.0\%      & \underline{64.0\%}     \\
Empowerment            & 36.0\%      & \underline{64.0\%}     & 42.4\%      & \underline{57.6\%}     & 22.8\%      & \underline{77.2\%}     & 47.6\%      & \underline{52.4\%}     \\
Overall                & 35.2\%      & \underline{64.8\%}     & 44.0\%      & \underline{56.0\%}     & 22.0\%      & \underline{78.0\%}     & 42.4\%      & \underline{57.6\%}     \\
\cmidrule(lr){2-3} \cmidrule(lr){4-5} \cmidrule(lr){6-7} \cmidrule(lr){8-9}
                      & NaiveRAG & \textbf{-Low} & NaiveRAG & \textbf{-Low} & NaiveRAG & \textbf{-Low} & NaiveRAG & \textbf{-Low} \\
\midrule
Comprehensiveness      & 36.0\%      & \underline{64.0\%}     & 43.2\%      & \underline{56.8\%}     & 19.2\%      & \underline{80.8\%}     & 36.0\%      & \underline{64.0\%}     \\
Diversity              & 28.0\%      & \underline{72.0\%}     & 39.6\%      & \underline{60.4\%}     & 13.6\%      & \underline{86.4\%}     & 33.2\%      & \underline{66.8\%}     \\
Empowerment            & 34.8\%      & \underline{65.2\%}     & 42.8\%      & \underline{57.2\%}     & 16.4\%      & \underline{83.6\%}     & 35.2\%      & \underline{64.8\%}     \\
Overall                & 34.8\%      & \underline{65.2\%}     & 43.6\%      & \underline{56.4\%}     & 18.8\%      & \underline{81.2\%}     & 35.2\%      & \underline{64.8\%}     \\

\cmidrule(lr){2-3} \cmidrule(lr){4-5} \cmidrule(lr){6-7} \cmidrule(lr){8-9}
                      & NaiveRAG & \textbf{-Origin} & NaiveRAG & \textbf{-Origin} & NaiveRAG & \textbf{-Origin} & NaiveRAG & \textbf{-Origin} \\
\midrule
Comprehensiveness      & 24.8\%      & \underline{75.2\%}     & 39.2\%      & \underline{60.8\%}     & 16.4\%      & \underline{83.6\%}     & 44.4\%      & \underline{55.6\%}     \\
Diversity              & 26.4\%      & \underline{73.6\%}     & 44.8\%      & \underline{55.2\%}     & 14.4\%      & \underline{85.6\%}     & 25.6\%      & \underline{74.4\%}     \\
Empowerment            & 32.0\%      & \underline{68.0\%}     & 43.2\%      & \underline{56.8\%}     & 17.2\%      & \underline{82.8\%}     & 45.2\%      & \underline{54.8\%}     \\
Overall                & 25.6\%      & \underline{74.4\%}     & 39.2\%      & \underline{60.8\%}     & 15.6\%      & \underline{84.4\%}     & 44.4\%      & \underline{55.6\%}     \\
\bottomrule
\end{tabular}
}
\end{table}

We also conduct ablation studies to evaluate the impact of our dual-level retrieval paradigm and the effectiveness of our graph-based text indexing in \model. The results are presented in Table~\ref{tab:ablation}.

\textbf{Effectiveness of Dual-level Retrieval Paradigm}.
We begin by analyzing the effects of low-level and high-level retrieval paradigms. We compare two ablated models—each omitting one module—against \model\ across four datasets. Here are our key observations for the different variants:

\begin{itemize}[leftmargin=*]
    \item \textbf{Low-level-only Retrieval}: The -High variant removes high-order retrieval, leading to a significant performance decline across nearly all datasets and metrics. This drop is mainly due to its emphasis on the specific information, which focuses excessively on entities and their immediate neighbors. While this approach enables deeper exploration of directly related entities, it struggles to gather information for complex queries that demand comprehensive insights.

    \item \textbf{High-level-only Retrieval}: The -Low variant prioritizes capturing a broader range of content by leveraging entity-wise relationships rather than focusing on specific entities. This approach offers a significant advantage in comprehensiveness, allowing it to gather more extensive and varied information. However, the trade-off is a reduced depth in examining specific entities, which can limit its ability to provide highly detailed insights. Consequently, this high-level-only retrieval method may struggle with tasks that require precise, detailed answers.

    \item \textbf{Hybrid Mode}: The hybrid mode, or the full version of \model, combines the strengths of both low-level and high-level retrieval methods. It retrieves a broader set of relationships while simultaneously conducting an in-depth exploration of specific entities. This dual-level approach ensures both breadth in the retrieval process and depth in the analysis, providing a comprehensive view of the data. As a result, \model\ achieves balanced performance across multiple dimensions.
    
\end{itemize}

\textbf{Semantic Graph Excels in RAG}.
We eliminated the use of original text in our retrieval process. Surprisingly, the resulting variant, -Origin, does not exhibit significant performance declines across all four datasets. In some cases, this variant even shows improvements (\textit{e.g.} in Agriculture and Mix). We attribute this phenomenon to the effective extraction of key information during the graph-based indexing process, which provides sufficient context for answering queries. Additionally, the original text often contains irrelevant information that can introduce noise in the response.

\subsection{Case Study (RQ3)}
To provide a clear comparison between baseline methods and our \model, we present specific case examples in Table~\ref{tab:case}, which includes responses to a machine learning question from both the competitive baseline, GraphRAG, and our \model\ framework. In this instance, \model\ outperforms in all evaluation dimensions assessed by the LLM judge, including comprehensiveness, diversity, empowerment, and overall quality. Our key observations are as follows:

\textbf{i) Comprehensiveness}. Notably, \model\ covers a broader range of machine learning metrics, showcasing its comprehensiveness and ability to effectively discover relevant information. This highlights the strength of our graph-based indexing paradigm, which excels in precise entity and relation extraction as well as LLM profiling.
\textbf{ii) Both Diversity and Empowerment}.
Furthermore, \model\ not only offers a more diverse array of information but also delivers more empowering content. This success is due to \model's hierarchical retrieval paradigm, which combines in-depth explorations of related entities through low-level retrieval to enhance empowerment with broader explorations via high-level retrieval to improve answer diversity. Together, these approaches capture a comprehensive global perspective of the knowledge domain, contributing to better RAG performance.

\begin{table}[t]
\centering
\caption{Case Study: Comparison Between \model\ and the Baseline Method GraphRAG.}
\vspace{-0.08in}
\label{tab:case}
\resizebox{\textwidth}{!}{
\begin{tabular}{|l|p{15cm}|}
\hline
\textbf{Query} & Which methods can normalize the feature values for effective machine learning? \\ \hline
\textbf{GraphRAG} & \textbf{Key Metrics for Evaluating Movie Recommendation Systems}:  
Evaluating the effectiveness of movie recommendation systems is crucial for ensuring user satisfaction and improving the overall performance of these systems. Several metrics are widely recognized for their ability to provide insights into how well a recommendation system performs. Below are the most informative metrics:  
1. \textbf{Precision} [...]  
2. \textbf{Recall} [...]  
3. \textbf{F1 Score} [...]  
4. \textbf{Mean Average Precision (MAP)} [...]  
5. \textbf{Root Mean Squared Error (RMSE)} [...]  
6. \textbf{User Satisfaction Metrics} [...]  
\textbf{Conclusion}: In summary, metrics such as precision, recall, F1 score, MAP, RMSE, and user satisfaction are critical for evaluating the performance of movie recommendation systems. Each metric provides unique insights that can help developers and researchers refine their algorithms and enhance user experiences. By focusing on these metrics, stakeholders can ensure that their recommendation systems effectively meet user preferences and expectations. \\ \hline
\textbf{LightRAG} & \textbf{Key Metrics for Evaluating Movie Recommendation Systems}:  
Evaluating movie recommendation systems involves multiple metrics that assess the effectiveness and accuracy of the recommendations provided to users. Below are some of the most informative metrics commonly used in this context, derived from the data and concepts surrounding recommendation systems.  
1. \textbf{Mean Average Precision at K (MAPK)} [...]  
2. \textbf{Precision and Recall} [...]  
3. \textbf{Root Mean Squared Error (RMSE) and Mean Squared Error (MSE)} [...]  
4. \textbf{Area Under the Curve (AUC)} [...]  
5. \textbf{F-Measure} [...]  
6. \textbf{User Engagement Metrics} [...]  
\textbf{Conclusion}: A comprehensive evaluation of movie recommendation systems employs a range of metrics that capture both the accuracy of the recommendations and user satisfaction. By effectively leveraging these metrics, system designers can optimize recommendation engines to deliver more personalized and relevant experiences. \\ \hline
\textbf{LLM Decision} & \textbf{Comprehensiveness}:  
"Winner": "Answer 2 (LightRAG)", 

"Explanation": "Answer 2 provides a broader array of metrics including MAPK, AUC, and user engagement metrics alongside precision, recall, RMSE, and F-Measure. This gives a fuller view of how recommendation systems can be evaluated. Answer 1, while detailed, is more focused on traditional metrics and lacks the inclusion of engagement metrics which are crucial for understanding user interaction." \newline

\textbf{Diversity}:  
"Winner": "Answer 2 (LightRAG)", 

"Explanation": "Answer 2 not only covers a wide variety of metrics but also includes nuanced explanations of how some metrics interrelate and differ from one another, like the inclusion of both RMSE and MSE, as well as the details behind AUC. In contrast, Answer 1 sticks primarily to standard metrics without much exploration of potential nuances." \newline

\textbf{Empowerment}:  
"Winner": "Answer 2 (LightRAG)", 

"Explanation": "Answer 2 empowers the reader more effectively by detailing how each metric functions and its importance in evaluating recommendation systems. By providing context such as the trade-offs between precision and recall and emphasizing user engagement metrics, it enables readers to make more informed judgments and understand the implications of different metrics. Answer 1 is more straightforward but lacks the depth of insight regarding why these metrics matter." \newline



\textbf{Overall Winner}:  
"Winner": "Answer 2 (LightRAG)",  

"Explanation": "While Answer 1 is more direct and systematic, Answer 2 excels in comprehensiveness, diversity, and empowerment. It provides a richer exploration of the topic, including insights into user engagement and nuanced differences between metrics. This depth and breadth make it more informative for readers seeking to thoroughly understand the evaluation of movie recommendation systems." \\ \hline

\end{tabular}
}
\vspace{-0.2in}
\end{table}

\vspace{-0.05in}
\subsection{Model Cost and Adaptability Analysis (RQ4)}
\vspace{-0.05in}

\begin{wrapfigure}{R}{0.45\textwidth}
    \vspace{-0.18in}
    \centering
    \caption{Comparison of Cost in Terms of Tokens and API Calls for GraphRAG and \model\ on the Legal Dataset.}
    \vspace{-0.1in}
    \label{tab:cost}
    \centering
    \scriptsize
    \setlength{\tabcolsep}{0.3mm}
    \begin{tabular}{|c|c|c|c|c|}
        \hline
        Phase & \multicolumn{2}{c|}{Retrieval Phase} & \multicolumn{2}{c|}{Incremental Text Update} \\ \hline
        Model & GraphRAG & Ours & GraphRAG & Ours \\ \hline
        \multirow{2}{*}{Tokens} & \multirow{2}{*}{$610 \times 1{,}000$} & \multirow{2}{*}{$< 100$} & $1{,}399 \times 2 \times 5{,}000$ & \multirow{2}{*}{$T_{\text{extract}}$} \\ 
        &&&$+ T_{\text{extract}}$& \\
        \hline
        API & \multirow{2}{*}{$\frac{610 \times 1{,}000}{C_{\text{max}}}$} & \multirow{2}{*}{1} & \multirow{2}{*}{$1{,}399 \times 2 + C_{\text{extract}}$} & \multirow{2}{*}{$C_{\text{extract}}$}\\
        Calls & &&&\\\hline
    \end{tabular}
    \vspace{-0.15in}
\end{wrapfigure}

We compare the cost of our \model\ with that of the top-performing baseline, GraphRAG, from two key perspectives. First, we examine the number of tokens and API calls during the indexing and retrieval processes. Second, we analyze these metrics in relation to handling data changes in dynamic environments. The results of this evaluation on the legal dataset are presented in Table~\ref{tab:cost}. In this context, $T_{\text{extract}}$ represents the token overhead for entity and relationship extraction, $C_{\text{max}}$ denotes the maximum number of tokens allowed per API call, and $C_{\text{extract}}$ indicates the number of API calls required for extraction.

In the retrieval phase, GraphRAG generates 1,399 communities, with 610 level-2 communities actively utilized for retrieval in this experiment. Each community report averages 1,000 tokens, resulting in a total token consumption of 610,000 tokens (610 communities $\times$ 1,000 tokens per community). Additionally, GraphRAG's requirement to traverse each community individually leads to hundreds of API calls, significantly increasing retrieval overhead. In contrast, \model\ optimizes this process by using fewer than 100 tokens for keyword generation and retrieval, requiring only a single API call for the entire process. This efficiency is achieved through our retrieval mechanism, which seamlessly integrates graph structures and vectorized representations for information retrieval, thereby eliminating the need to process large volumes of information upfront.

In the incremental data update phase, designed to address changes in dynamic real-world scenarios, both models exhibit similar overhead for entity and relationship extraction. However, GraphRAG shows significant inefficiency in managing newly added data. When a new dataset of the same size as the legal dataset is introduced, GraphRAG must dismantle its existing community structure to incorporate new entities and relationships, followed by complete regeneration. This process incurs a substantial token cost of approximately 5,000 tokens per community report. Given 1,399 communities, GraphRAG would require around 1,399 × 2 × 5,000 tokens to reconstruct both the original and new community reports—an exorbitant expense that underscores its inefficiency. In contrast, \model\ seamlessly integrates newly extracted entities and relationships into the existing graph without the need for full reconstruction. This approach results in significantly lower overhead during incremental updates, demonstrating its superior efficiency and cost-effectiveness.

\vspace{-0.1in}
\section{Related Work}
\label{sec:relate}

\vspace{-0.1in}
\subsection{Retrieval-Augmented Generation with LLMs}\vspace{-0.05in}
Retrieval-Augmented Generation (RAG) systems enhance LLM inputs by retrieving relevant information from external sources, grounding responses in factual, domain-specific knowledge \cite{ram2023context, fan2024survey}. Current RAG approaches \cite{hyde, gao2023retrieval, chan2024rq, yu2024rankrag} typically embed queries in a vector space to find the nearest context vectors. However, many of these methods rely on fragmented text chunks and only retrieve the top-k contexts, limiting their ability to capture comprehensive global information needed for effective responses.

Although recent studies \cite{edge2024local} have explored using graph structures for knowledge representation, two key limitations persist. First, these approaches often lack the capability for dynamic updates and expansions of the knowledge graph, making it difficult to incorporate new information effectively. In contrast, our proposed model, \model, addresses these challenges by enabling the RAG system to quickly adapt to new information, ensuring the model's timeliness and accuracy. Additionally, existing methods often rely on brute-force searches for each generated community, which are inefficient for large-scale queries. Our \model\ framework overcomes this limitation by facilitating rapid retrieval of relevant information from the graph through our proposed dual-level retrieval paradigm, significantly enhancing both retrieval efficiency and response speed.

\vspace{-0.1in}
\subsection{Large Language Model for Graphs}\vspace{-0.05in}
Graphs are a powerful framework for representing complex relationships and find applications in numerous fields. As Large Language Models (LLMs) continue to evolve, researchers have increasingly focused on enhancing their capability to interpret graph-structured data. This body of work can be divided into three primary categories: i) \textbf{GNNs as Prefix} where Graph Neural Networks (GNNs) are utilized as the initial processing layer for graph data, generating structure-aware tokens that LLMs can use during inference. Notable examples include GraphGPT~\cite{tang2024graphgpt} and LLaGA~\cite{chenllaga}. ii) \textbf{LLMs as Prefix} involves LLMs processing graph data enriched with textual information to produce node embeddings or labels, ultimately refining the training process for GNNs, as demonstrated in systems like GALM~\cite{xie2023graph} and OFA~\cite{liuone}. iii) \textbf{LLMs-Graphs Integration} focuses on achieving a seamless interaction between LLMs and graph data, employing techniques such as fusion training and GNN alignment, and developing LLM-based agents capable of engaging with graph information directly~\cite{li2023grenade,brannon2023congrat}.
\vspace{-0.1in}
\section{Conclusion}\vspace{-0.1in}
\label{sec:conclusion}

This work introduces an advancement in Retrieval-Augmented Generation (RAG) through the integration of a graph-based indexing approach that enhances both efficiency and comprehension in information retrieval. \model\ utilizes a comprehensive knowledge graph to facilitate rapid and relevant document retrieval, enabling a deeper understanding of complex queries. Its dual-level retrieval paradigm allows for the extraction of both specific and abstract information, catering to diverse user needs. Furthermore, \model's seamless incremental update capability ensures that the system remains current and responsive to new information, thereby maintaining its effectiveness over time. Overall, \model\ excels in both efficiency and effectiveness, significantly improving the speed and quality of information retrieval and generation while reducing costs for LLM inference.

\clearpage

\bibliography{refs}
\bibliographystyle{iclr2025_conference}

\clearpage
\section{Appendix}
\label{app:appendix}
In this section, we elaborate on the methodologies and experimental settings used in the \model\ framework. It describes the specific steps for extracting entities and relationships from documents, detailing how large language models (LLMs) are utilized for this purpose. The section also specifies the prompt templates and configurations used in LLM operations, ensuring clarity in the experimental setup. Additionally, it outlines the evaluation criteria and dimensions used to assess the performance of \model\ against baselines from various dimensions.

\subsection{Experimental Data Details}

\begin{table}[h]
\caption{Statistical information of the datasets.}
\label{dataset}
\centering
\begin{tabular}{lcccc}
\toprule
\textbf{Statistics}  & \textbf{Agriculture} & \textbf{CS} & \textbf{Legal} & \textbf{Mix} \\
\midrule
Total Documents & 12  & 10  & 94  & 61  \\
Total Tokens    & 2,017,886 & 2,306,535 & 5,081,069 & 619,009 \\
\bottomrule
\end{tabular}
\end{table}

Table~\ref{dataset} presents statistical information for four datasets: Agriculture, CS, Legal, and Mix. The Agriculture dataset consists of 12 documents totaling 2,017,886 tokens, while the CS dataset contains 10 documents with 2,306,535 tokens. The Legal dataset is the largest, comprising 94 documents and 5,081,069 tokens. Lastly, the Mix dataset includes 61 documents with a total of 619,009 tokens.

\vspace{-0.1in}
\subsection{Case Example of Retrieval-Augmented Generation in \model.}
\label{app:retrieval_and_generation_examples}
\begin{figure}[h]
\centering
\includegraphics[width=1\textwidth]{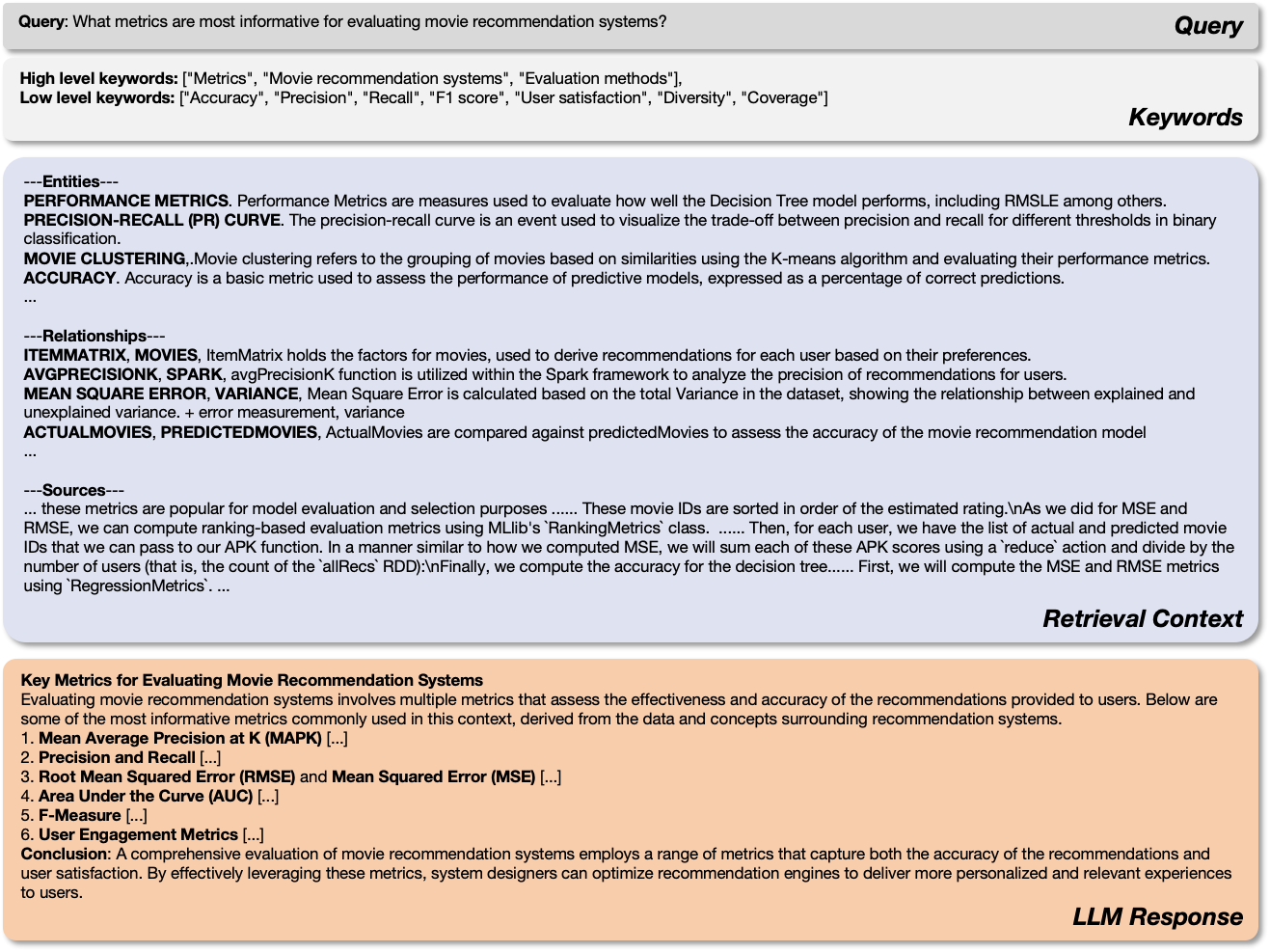}
\caption{A retrieval and generation example.}
\label{fig:retrieval_and_generation_example}
\end{figure}
In Figure \ref{fig:retrieval_and_generation_example}, we illustrate the retrieve-and-generate process. When presented with the query, ``What metrics are most informative for evaluating movie recommendation systems?'', the LLM first extracts both low-level and high-level keywords. These keywords guide the dual-level retrieval process on the generated knowledge graph, targeting relevant entities and relationships. The retrieved information is organized into three components: entities, relationships, and corresponding text chunks. This structured data is then fed into the LLM, enabling it to generate a comprehensive answer to the query.

\vspace{-0.1in}
\subsection{Overview of the Prompts Used in \model}
\subsubsection{Prompts for Graph Generation}
\label{app:graph_construct_prompt}

\begin{figure}[h]
\centering
\includegraphics[width=1\textwidth]{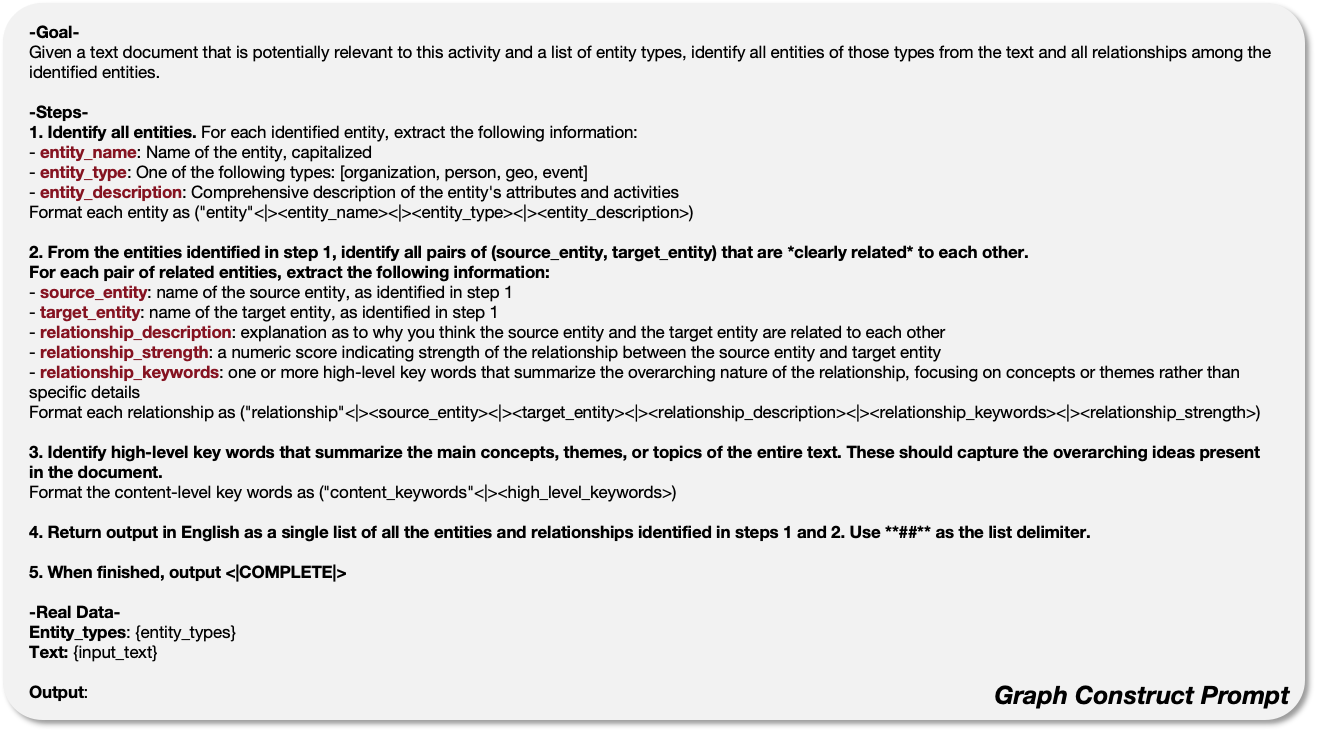}
\caption{Prompts for Graph Generation}
\label{fig:graph_construct_prompt}
\end{figure}

The graph construction prompt outlined in Figure~\ref{fig:graph_construct_prompt} is designed to extract and structure entity-relationship information from a text document based on specified entity types. The process begins by identifying entities and categorizing them into types such as organization, person, location, and event. It then provides detailed descriptions of their attributes and activities. Next, the prompt identifies relationships between these entities, offering explanations, assigning strength scores, and summarizing the relationships using high-level keywords.

\subsubsection{Prompts for Query Generation}
\label{app:graph_construct_prompt}

\begin{figure}[h]
\centering
\includegraphics[width=1\textwidth]{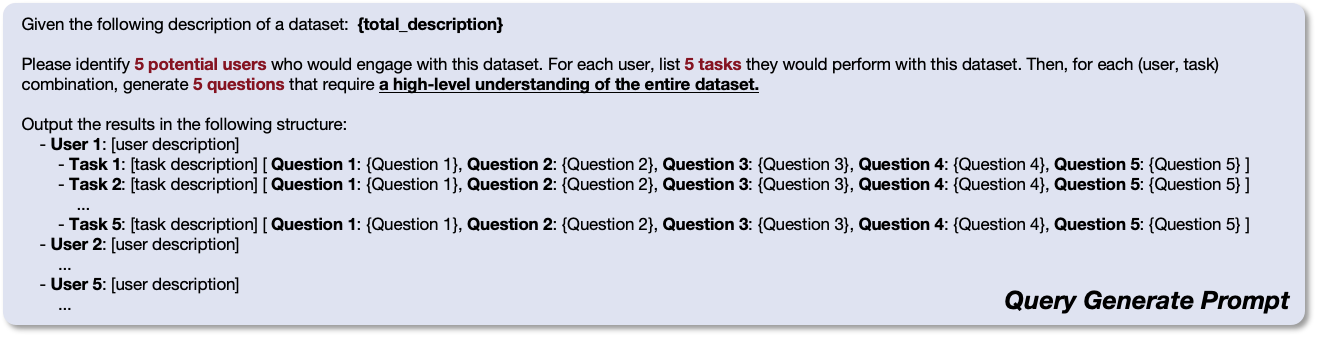}
\caption{Prompts for Query Generation}
\label{fig:query_generate_prompt}
\end{figure}

In Figure~\ref{fig:query_generate_prompt}, the query generation prompt outlines a framework for identifying potential user roles (e.g., data scientist, finance analyst, and product manager) and their objectives for generating queries based on a specified dataset description. The prompt explains how to define five distinct users who would benefit from interacting with the dataset. For each user, it specifies five key tasks they would perform while working with the dataset. Additionally, for each (user, task) combination, five high-level questions are posed to ensure a thorough understanding of the dataset.

\subsubsection{Prompts for Keyword Extraction}
\label{app:keywords_generate_prompt}

\begin{figure}[h]
\centering
\includegraphics[width=1\textwidth]{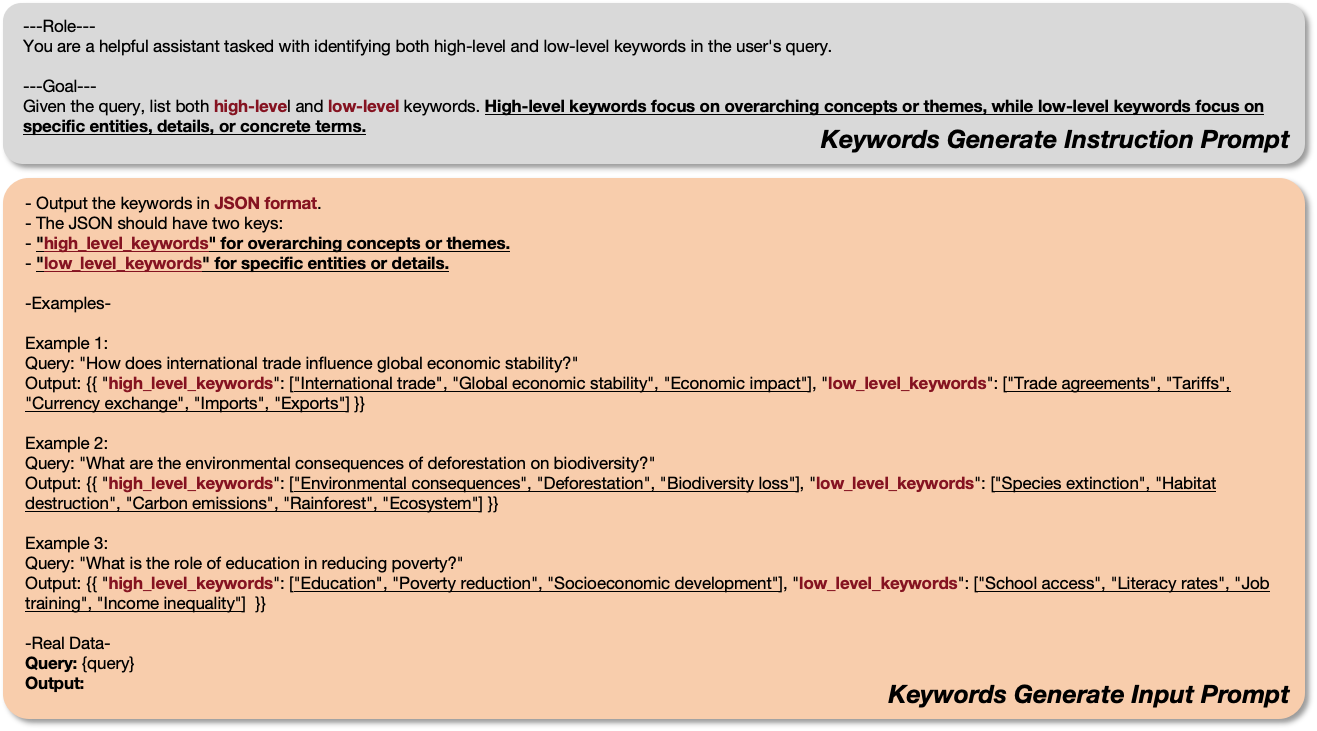}
\caption{Prompts for Keyword Extraction}
\label{fig:keywords_generate_prompt}
\end{figure}

In Figure~\ref{fig:keywords_generate_prompt}, the prompt describes a method for extracting keywords from a user's query, distinguishing between high-level and low-level keywords. High-level keywords represent broad concepts or themes, while low-level keywords focus on specific entities and details. The extracted keywords are returned in JSON format, organized into two fields: ``high\_level\_keywords'' for overarching ideas and ``low\_level\_keywords'' for specific details.

\subsubsection{Prompts for RAG Evaluation}
\label{app:evaluation_prompt}

\begin{figure}[h]
\centering
\includegraphics[width=1\textwidth]{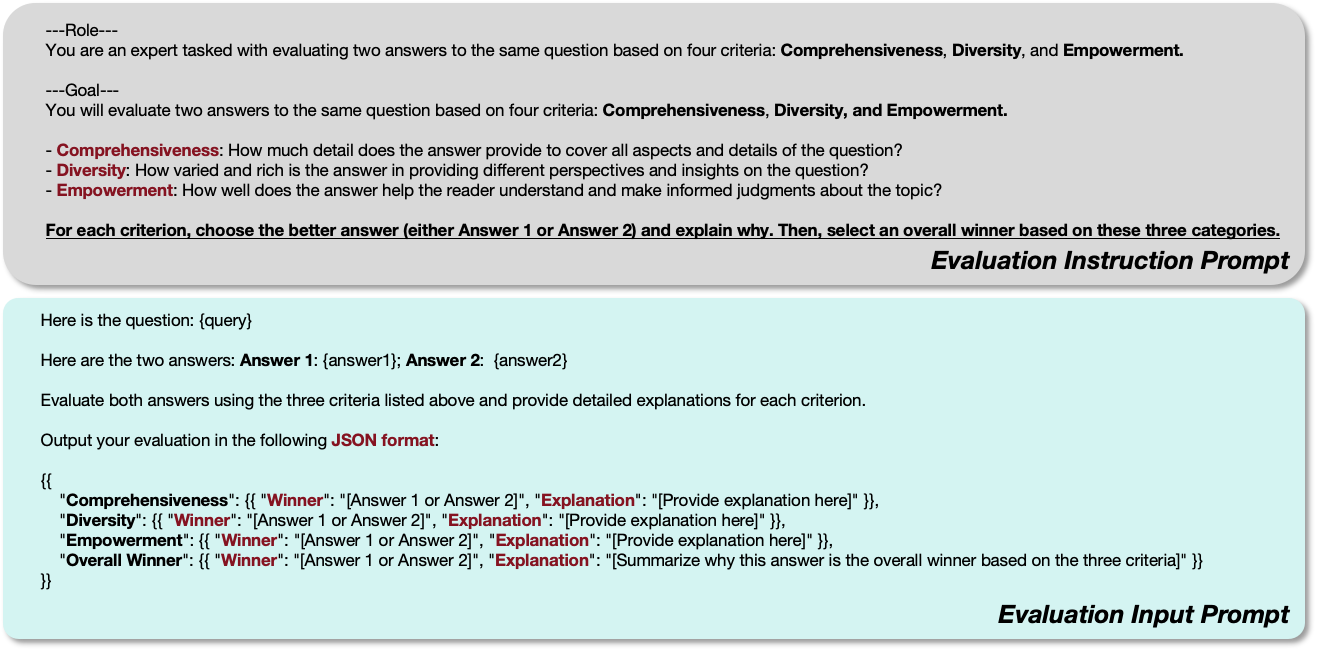}
\caption{Prompts for RAG Evaluation}
\label{fig:evaluation_prompt}
\end{figure}

The evaluation prompt is illustrated in Figure~\ref{fig:evaluation_prompt}. It introduces a comprehensive evaluation framework for comparing two answers to the same question based on three key criteria: Comprehensiveness, Diversity, and Empowerment. Its purpose is to guide the LLM through the process of selecting the better answer for each criterion, followed by an overall assessment. For each of the three criteria, the LLM must identify which answer performs better and provide a rationale for its choice. Ultimately, an overall winner is determined based on performance across all three dimensions, accompanied by a detailed summary that justifies the decision. The evaluation is structured in JSON format, ensuring clarity and consistency, and facilitating a systematic comparison between the two answers.

\vspace{-0.1in}
\subsection{Case Study: Comparison Between \model\ and the Baseline NaiveRAG.}

\begin{table}[t]
\centering
\caption{Case Study: Comparison Between \model\ and the Baseline NaiveRAG.}
\vspace{-0.08in}
\label{tab:case_2}
\resizebox{\textwidth}{!}{
\begin{tabular}{|l|p{15cm}|}
\hline
\textbf{Query} & How do indigenous perspectives on ownership and collaboration influence corporate mergers in countries like Canada and Australia? \\ \hline
\textbf{NaiveRAG} & Indigenous perspectives on ownership and collaboration play a significant role in shaping corporate mergers in countries like Canada and Australia, particularly where resources and land rights intersect with Indigenous sovereignty and cultural values. These perspectives emphasize community engagement, respect for traditional land use, and a collaborative approach to resource management, influencing corporate strategies and outcomes.

\textellipsis

Both nations illustrate that prioritizing Indigenous perspectives leads to more sustainable and equitable outcomes. Acknowledging Indigenous rights and fostering collaborative relationships can reduce conflicts and enhance corporate reputations. Ultimately, integrating Indigenous perspectives is a pathway to better business practices and community relations.

 \\ \hline
\textbf{LightRAG} & \textbf{Indigenous Perspectives on Ownership and Collaboration in Corporate Mergers}

Indigenous perspectives significantly influence corporate mergers in countries like Canada and Australia. These views emphasize communal rights to land and resources, spiritual connections to the environment, and the importance of collaboration over competition.

\textbf{Cultural Significance of Land Ownership} [...]

\textbf{The Role of Collaboration} [...]

\textbf{Legal and Regulatory Frameworks} [...] \newline

\textbf{Conclusion}

Indigenous perspectives shape corporate mergers by emphasizing communal ownership and collaborative partnerships. Corporations must adopt respectful, inclusive practices that honor Indigenous cultural values and rights, paving the way for sustainable, mutually beneficial outcomes.

 \\ \hline
\textbf{LLM Decision} & \textbf{Comprehensiveness}:  
"Winner": "Answer 2 (LightRAG)", 

"Explanation": "Answer 2 provides a thorough exploration of the indigenous perspectives by discussing cultural significance, collaboration, and legal frameworks, including specific examples and detailed insights. In contrast, Answer 1, while informative, lacks the same depth in analyzing the various dimensions of the indigenous perspectives on ownership and collaboration." \newline

\textbf{Diversity}:  
"Winner": "Answer 2 (LightRAG)", 

"Explanation": "Answer 2 presents a wider array of perspectives, focusing on the communal aspect of land ownership, spiritual connections, and practical examples of collaboration. It covers more ground by contrasting Indigenous views with Western notions, whereas Answer 1 primarily focuses on corporate strategies and outcomes with limited perspective." \newline

\textbf{Empowerment}:  
"Winner": "Answer 2 (LightRAG)", 

"Explanation": "Answer 2 equips the reader with nuanced understanding and actionable insights by highlighting the importance of collaboration and the legal frameworks involved. It empowers corporations by illustrating how adopting an inclusive approach can facilitate better outcomes. Answer 1, while practical, does not emphasize the moral or ethical implications as strongly." \newline

\textbf{Overall Winner}:  
"Winner": "Answer 2 (LightRAG)",  

"Explanation": "Answer 2 excels overall due to its comprehensive exploration, diversity of perspectives, and empowerment of the reader with actionable insights about indigenous perspectives and collaboration in corporate mergers. Although Answer 1 is more direct, the depth and breadth of Answer 2 make it the stronger response." \\ \hline

\end{tabular}
}
\end{table}

To further illustrate \model's superiority over baseline models in terms of comprehensiveness, empowerment, and diversity, we present a case study comparing \model\ and NaiveRAG in Table~\ref{tab:case_2}. This study addresses a question regarding indigenous perspectives in the context of corporate mergers. Notably, \model\ offers a more in-depth exploration of key themes related to indigenous perspectives, such as cultural significance, collaboration, and legal frameworks, supported by specific and illustrative examples. In contrast, while NaiveRAG provides informative responses, it lacks the depth needed to thoroughly examine the various dimensions of indigenous ownership and collaboration. The dual-level retrieval process employed by \model\ enables a more comprehensive investigation of specific entities and their interrelationships, facilitating extensive searches that effectively capture overarching themes and complexities within the topic.

\end{document}